\documentstyle[12pt]{article}

\def\Jou#1#2#3#4{{#1} {\bf #2} (#4) #3}

\def\PLB{{Phys. Lett.}  B}

\def\PRD{{Phys. Rev.} D}

\def\ZPA{{Z. Phys.} A}
\def\ZPC{{Z. Phys.} C}

\newfam\BMath
\font\BMathL=cmmib10 
\font\BMathl=cmmib7
\font\BMathm=cmmib5
\textfont\BMath=\BMathL \scriptfont\BMath=\BMathl
\scriptscriptfont\BMath=\BMathm

\def\Q{{\fam\BMath q}}
\def\K{{\fam\BMath k}}

\def\a{\alpha}

\def\c{\chi}

\def\f{\phi}

\def\j{\psi}

\def\cm{{\cal M}}

\def\dd{\mbox{d}}

\def\exp{\mbox{\rm exp}}

\def\ra{\rightarrow}

\def\lra{\longrightarrow}

\def\dxb{[\dd x]}

\def\dkb{[\dd^2 \Kp]}

\def\Kp{{\K_\perp}}
\def\Kpd{{\K'_\perp}}

\def\be{\begin{equation}}
\def\ee{\end{equation}}
\def\bea{\begin{eqnarray}}
\def\eea{\end{eqnarray}}
\def\bfig{\begin{figure}}
\def\efig{\end{figure}}
\def\eref#1{Eq.~(\ref{#1})}

\def\bfi{\begin{figure}}
\def\efi{\end{figure}}

\newcommand{\ncom}{\newcommand}
\ncom{\lan}{\langle}
\ncom{\ran}{\rangle}
\ncom\fx{\!\!\!\!}

\begin{document}

\begin{flushright}
\footnotesize \sffamily WU-B 97-27
\end{flushright}

{\Large\bf Colour Octet Contribution in Exclusive
Charmonium Decay into Baryon-Antibaryon$^*$ }\\[2mm]
{\it P. Kroll and S.M.H. Wong$^\dagger$}\\[2mm]
{\small Fachbereich Physik, Universit\"at Wuppertal, 
        D-42097 Wuppertal, Germany}\\[2mm]
{\bf Abstract}\\

It is argued that colour octet contribution in exclusive
charmonium decays is very important even though the same
infrared divergence found in P-wave colour singlet
inclusive decay is absent. This comes about because of
the suppression at the level of the wavefunction of P-wave 
quarkonia. Using the modified hard scattering approach of 
Brodsky, Lepage, Botts, Li and Sterman, we use charmonium 
decay into baryon-antibaryon pair as an example and show
that the colour singlet contribution alone is clearly 
insufficient to explain the experimental decay widths.

\normalsize

\section{Introduction}

It is known that in inclusive decay of P-wave charmonium into
light hadrons, there is an infrared (IR) divergence$^1$ in the
reaction $c\bar c \lra g q\bar q$ which is the leading 
inclusive contribution for $\chi_1$ but next-to-leading for 
$\chi_0$ and $\chi_2$. This IR divergence reveals itself in the
limit that the gluon becomes soft and the heavy $c\bar c$ are
allowed to go on the mass-shell. In view of the divergence
and the $c\bar c$ is a bounded system, one is obliged to keep
them off-shell by an amount of the binding energy $\varepsilon$.
The resulting inclusive decay width has a logarithmic dependence
on this binding energy $\Gamma^{incl} \propto \ln \varepsilon$.
However, it was shown in Ref. 2 that, in fact, 
this infrared divergence is there only due to the neglect of the 
next higher Fock state of the P-wave charmonium, the so-called
colour octet component, which becomes degenerate with the valence 
$c\bar c$ singlet state in the already mentioned decay process 
in the soft gluon limit. 

In exclusive P-wave charmonium decay, this same IR divergence
is not present because it only happens in a small region in
momentum space and the decay products are again bound states. 
Therefore a priori, there is no good reason to introduce the 
colour octet. Indeed, the exclusive decay widths for $\c_J$ 
have been calculated for a number of decay modes using the 
colour singlet valence Fock state alone and claims have been
made that these calculations can account for the corresponding
experimental partial widths. In the following, using the example of
$\c_J \lra N\bar N$, we shall argue that, in fact, there is a very 
good reason to include the colour octet component in exclusive 
decay even in the absence of the IR divergence as found in the 
calculation of the inclusive width and point out a fallacy 
of some of the previous works. We show 
that within a self-consistent perturbative scheme, the 
modified hard scattering approach, colour singlet contribution 
alone is insufficient to account for the experimental data. 
Thus, based on the theoretical argument to be presented, 
colour octet is required.

\rule{60mm}{0.4mm}

\vspace{1mm} \noindent
{\rm
$^*$ Work supported in part by European TMR network contract 
no. ERB-4061-PI95 \\
$^\dagger$ Talk presented by
}

\section{Colour Singlet vs Colour Octet}

The decay of the $c\bar c$ system through annihilation is a
short distance process with an annihilation size, $L$, of the 
order of the inverse charmonium mass $L\; \sim 1/M_{\c_J} << 1$. 
So for S-wave decay, one needs essentially $\j_s(L\sim 0)$,
the wavefunction at the origin. For a P-wave system, the 
wavefunction at the origin is zero, so instead one uses
$L \j_p'(L\sim 0)$, the derivative of the wavefunction
at the origin weighed by the annihilation size. In momentum
space, the contributions from the wavefunctions alone
will provide a P-wave to S-wave ratio of 
$k \j_p(k)/M_{\c_J} \j_s(k)$ so P-wave is suppressed in 
comparison to S-wave by $1/M_{\c_J}$. This suppression is 
very important, as we shall see presently, for whether 
the colour octet component should be included. 

In the hard scattering picture of Brodsky and Lepage,
the decay probability amplitude is given by a convolution
of the distribution amplitudes and the hard scattering
perturbative part
\be \cm_{\c_J\lra N\bar N} \; \sim \; f_N \f_N \otimes
    f_{\bar N} \f_{\bar N} \otimes f_{\c_J} \f_{\c_J} 
    \otimes T_H  \; .
\label{eq:conv_hsp}
\ee
The only dimensional quantities in the amplitude are the
decay constants and some power of the large scale of the 
process in question, namely the charmonium mass, hidden in 
the hard part $T_H$. All these must together make up
the right dimension of the probability amplitude which 
has one mass dimension because it is related to the decay
width through
\be \Gamma \sim \frac{1}{M_{\c_J}} |\cm |^2 \; .
\ee
The decay constants in \eref{eq:conv_hsp} all have mass 
dimension two for both colour singlet and octet contribution to 
the width because both the wavefunctions of the baryon and the 
colour octet component are three-particle wavefunctions while
the colour singlet component is a P-wave two-particle
wavefunction. Using the right power of the charmonium mass from
the hard part $T_H$ to make up for the right mass dimension of 
the amplitude gives 
\bea \cm^{(1)}_{\c_J \lra N\bar N}  & \sim & \fx M_{\c_J} 
     \left (\frac{f_B}{M_{\c_J}^2} \right )^2
     \left (\frac{f^{(1)}_{\c_J}}{M_{\c_J}^2} \right ) 
     \sim \frac{1}{M_{\c_J}^5}                                \\
     \cm^{(8)}_{\c_J \lra N\bar N}  & \sim & \fx M_{\c_J} 
     \left (\frac{f_B}{M_{\c_J}^2} \right )^2
     \left (\frac{f^{(8)}_{\c_J}}{M_{\c_J}^2} \right )
     \sim \frac{1}{M_{\c_J}^5}                                \; .
\eea
As one can see, both the octet and singlet contribution 
depend on the same power of $M_{\c_J}$. The octet contribution 
is not negligible as far as the large scale $M_{\c_J}$ is 
concerned. This is the case only because, as we saw above, 
P-wave charmonium is weighed down by the annihilation 
size $L \sim 1/M_{\c_J}$ in 
the amplitude. Had we been dealing with the S-wave 
charmonium $J/\j$ instead, the singlet amplitude would go as 
$1/M_{\c_J}^4$ and the colour octet can be neglected in 
comparison with the colour singlet valence component.
Returning to the $\c_J$ decay, possible suppression from
the $\a_s$ dependence is absence. The singlet and octet 
contribution are only differed by $\sqrt{\a_s}$ for typical graphs
which is not small. Therefore there is no good reason to 
neglect the colour octet from exclusive decay.

\section{Colour Singlet Contribution within the Modified \\ 
Hard Scattering Approach}

In the modified hard scattering picture, the decay probability
amplitude is given by 
\be \cm_{\c_J \lra N\bar N} \sim \j_B \otimes \j_{\bar B} \otimes 
    \j_{\c_J} \otimes T_H \otimes \exp (-S)    \; .
\ee
a convolution of the wavefunctions, hard part and Sudakov factor.
In the valence colour singlet contribution, the part of the 
convolution that can potentially render the result ambiguous 
is the nucleon wavefunction. The reason being that there are many 
nucleon model wavefunctions available that one can use, for example 
see Ref. 3, for this calculation. However, none of them can 
explain data on nucleon form factor at low $Q^2 < 50$ GeV$^2$,
which is the relevant energy range of interest in our present 
problem. As shown in Ref. 4, these models yield nucleon form factor 
results that are well below the experimental data. As a result, 
one cannot possibly hope to use these models to calculate
the $\c_J$ partial decay width. In view of the smallness
of the perturbative contribution, one is forced to leave
the perturbative region and find non-perturbative 
contribution which can account for the data. The full 
expression for the nucleon form factor can be written as
\be F_1 (Q^2) = \sum^\infty_{n=3} \int \dxb \dkb 
    \sum^n_{j \in \{q, \; \bar q\} } \; 
    e_j \; \j^{(n)}(x,\Kp) \; \j^{(n)*} (x,\Kpd)
\label{eq:full_f2}
\ee
where $\Kpd = \Kp + (1-x_i) \Q\;$ for the struck quark and
$\Kpd = \Kp - x_j \Q\;$ for the spectator quarks,
the second sum is over the parton index of the quark and 
antiquark constituents and we have only put a representative
$x$ and $\Kp$ to stand for all the momentum fractions and 
internal momenta of the constituents in the argument of the 
$n$-constituent nucleon wavefunction $\j^{(n)}$. When hit by 
a highly virtual photon, one knows that the perturbative contribution 
comes from the region where the internal momenta of either one of 
the wavefunction in \eref{eq:full_f2} are hard while those in the 
other remain soft. Since this is not the dominant contribution, 
one looks at the region where the internal momenta in both 
wavefunctions are soft. The reason being that bounded hadrons 
tend to have small internal momentum so if the incoming photon
momentum is not so hard, as in the low $Q^2$ but $Q^2>>$ MeV$^2$ 
region, the overlap of both wavefunctions with small internal 
momentum should dominate. This can happen only if the struck 
quark takes up almost all the momentum of the nucleon so that 
$(1-x_i) \Q$ and $x_j \Q$ all become soft. This is most unlikely 
if many constituent partons are there to share the nucleon momentum.
So one can take only the valence Fock state as the dominant
contribution
\be F_1 (Q^2) \simeq \int \dxb \dkb \sum^3_{j=1} \; 
    e_j \; \j^{(3)}(x,\Kp) \; \j^{(3)*} (x,\Kpd)  \; .
\label{eq:val_f2}
\ee 
This simplication enables one to fit this to the $F_1$ 
data, GRV valence quark distribution and other data like 
$\Gamma(J/\j \lra N\bar N)$ to obtain a model wavefunction
for the nucleon$^5$ valid in the low $Q^2$ range of interest.  
The other remaining colour singlet $\c_J$ wavefunction is more
or less standard (see for example Ref. 6). The hard part 
$T_H$ in the MHSP comes from 4 graphs up to permutation
of the light quark-antiquark lines$^7$. Here we do not give 
the full expression but just to point out that in the MHSP, 
the probability amplitude has the following dependence on
$\a_s$
\be \cm^{(1)}_{\c \lra N\bar N} \; \sim \; \a_s(t_1) \; \a_s(t_2) \; 
    \a_s(t_3)
\ee
where the renormalization scales $t_i$, $i=1,2,3$ are determined
by the largest scale in the neighbourhood of the vertices that carry
the particular $\a_s$. The scales could be the virtualities of the 
neighbouring propagators or the inverse of the smallest transverse 
size squared between the quarks (antiquarks) in the baryon 
(antibaryon). So the renormalization scales are determined 
dynamically by the process which is a typical hallmark of the
MHSP. The product of $\a_s$'s are therefore part of the 
integrand of the convolution. One cannot, as done in some earlier 
work, take this product outside the integrand and assign some
constant effective value for the coupling. In our opinion, this 
procedure is rather unconvincing because $\cm^{(1)} \sim \a_s^3$ 
and therefore $\Gamma^{(1)} \sim \a_s^6$. Changing $\a_s$ from a 
value of $0.3$ to $0.5$ will make a factor of $20$ difference in 
the width so provided the singlet contribution is of reasonable
size, one can always choose a value such that the colour singlet 
alone can account for the experimental width. 

As we argued in the previous section, colour octet is not 
suppressed for P-wave decay with respect to the singlet 
contribution therefore both should be needed to account
for the experimental data. In table \ref{tab:sing_nucl}, we listed 
the singlet contributions to $\c_J \lra p\bar p$ partial width. 
$\c_0$ is forbidden in this decay mode by angular momentum
conservation. One can see that the singlet contribution
can only account for $10\%$ of $\c_1$ and $24\%$ of $\c_2$
of the experimental partial width. So in agreement with our
theoretical argument presented in the previous section, 
colour octet is needed in exclusive P-wave charmonium decay even
in the absence of the same infrared divergence as found in 
the colour singlet inclusive decay width, to cancel which the
colour octet was introduced. 

\begin{table}
\begin{center}
\begin{tabular}{||c||c|c||}
\hline
  $J$ & $\Gamma^{(1)}(\c_J\ra p\bar p)$ [eV]
      & PDG$^8$ [eV] \\
\hline \hline
 1  & \ 7.49  & \  75.68   \\ \hline
 2  &  46.97  &   200.00   \\ \hline
\end{tabular}
\end{center}
\caption{Colour singlet contributions$^7$ to $\c_J$ decay into 
$p\bar p$.}
\label{tab:sing_nucl}
\end{table}

Our theoretical argument and conclusion are however much more 
general than just for charmonium decay alone. One could apply these
equally to bottomium as well as to P-wave quarkonium production. 
In these processes, colour octet should be required. One can
however venture even further by making the suggestion that
whenever a P-wave heavy quark-antiquark pair, which could be a
part of the constituents of some component of a hadronic 
wavefunction, is needed, one will also have the need of the 
component of the corresponding colour octet accompanying 
the same set of other constituents as well.


\noindent
{ \\
1. \ R. Barbieri and R. Gatto, \Jou{\PLB}{61}{465}{1976}. \\ 
2. \ G.T. Bodwin, E. Braaten and G.P. Lepage, \Jou{\PRD}{46}{1914}{1992},
   {\bf 51} (1995) \\
\null \hspace{0.50cm} 1125. \\
3. \ M. Bergmann and N.G. Stefanis, \Jou{\PRD}{48}{R2990}{1993}. \\
4. \ J. Bolz, R. Jacob, P. Kroll, M. Bergmann and N.G. Stefanis,
\Jou{\ZPC}{66}{}{1995} \\
\null \hspace{0.50cm} 267. \\
5. \ J. Bolz and P. Kroll, \Jou{\ZPA}{356}{327}{1996}. \\
6. \ J. Bolz and P. Kroll and G. Schuler, \Jou{\PLB}{392}{198}{1997}. \\
7. \ P. Kroll and S.M.H. Wong, work in progress. \\
8. \ R.M. Barnett et al, \Jou{\PRD}{54}{1}{1996}. 
}

\end{document}